\documentclass[twocolumn,superscriptaddress,floatfix,prl,showpacs]{revtex4-1}
\usepackage{graphicx}
\usepackage{amsmath}
\usepackage{amssymb}
\usepackage{color}
\usepackage{braket}
\usepackage{dsfont}
\usepackage{lineno}

\begin{document}
\title{Quantum Phase Transitions in the Spin-Boson model: MonteCarlo Method vs Variational Approach {\it a la} Feynman}

\author{G. De Filippis}
\affiliation{SPIN-CNR and Dip. di Fisica - Universit\`a di Napoli Federico II - I-80126 Napoli, Italy}
\affiliation{INFN, Sezione di Napoli - Complesso Universitario di Monte S. Angelo - I-80126 Napoli, Italy}

\author{A. de Candia}
\affiliation{SPIN-CNR and Dip. di Fisica - Universit\`a di Napoli Federico II - I-80126 Napoli, Italy}
\affiliation{INFN, Sezione di Napoli - Complesso Universitario di Monte S. Angelo - I-80126 Napoli, Italy}

\author{L.~M.~Cangemi}
\affiliation{SPIN-CNR and Dip. di Fisica - Universit\`a di Napoli Federico II - I-80126 Napoli, Italy}

\author{M. Sassetti}
\affiliation{Dipartimento di Fisica, Universit\`a di Genova, I-16146 Genova, Italy}
\affiliation{SPIN-CNR, I-16146 Genova, Italy}

\author{R. Fazio}
\affiliation{SPIN-CNR and Dip. di Fisica - Universit\`a di Napoli Federico II - I-80126 Napoli, Italy}
\affiliation{ICTP, Strada Costiera 11, I-34151 Trieste, Italy}
\affiliation{NEST, Istituto Nanoscienze-CNR, I-56126 Pisa, Italy}

\author{V. Cataudella}
\affiliation{SPIN-CNR and Dip. di Fisica - Universit\`a di Napoli Federico II - I-80126 Napoli, Italy}
\affiliation{INFN, Sezione di Napoli - Complesso Universitario di Monte S. Angelo - I-80126 Napoli, Italy} 

\begin{abstract}
  The effectiveness of the variational approach a la Feynman is proved in the spin-boson model, i.e. the simplest realization of the Caldeira-Leggett model able to reveal the quantum phase transition from delocalized to localized states and the quantum dissipation and decoherence effects induced by a heat bath. After exactly eliminating the bath degrees of freedom, we propose a trial, non local in time, interaction between the spin and itself simulating the coupling of the two level system with the bosonic bath. It stems from an Hamiltonian where the spin is linearly coupled to a finite number of harmonic oscillators whose frequencies and coupling strengths are variationally determined. We show that a very limited number of these fictitious modes is enough to get a remarkable agreement, up to very low temperatures, with the data obtained by using an approximation-free Monte Carlo approach, predicting: 1) in the Ohmic regime, a Beretzinski-Thouless-Kosterlitz quantum phase transition exhibiting the typical universal jump at the critical value; 2) in the sub-Ohmic regime ($s \leq 0.5$), mean field quantum phase transitions, with logarithmic corrections for $s=0.5$.
\end{abstract}
\maketitle

The spin-boson model is a paradigmatic minimal model used for describing the quantum phase transition (QPT) from delocalized to localized states induced by the environment \cite{leggett,weiss,hur}. It also plays a significant role in understanding the relaxation processes, in particular the dissipation and decoherence effects, in quantum systems \cite{Breuer,revmod}. The model consists of a two-level system, i.e. a single quantum qubit or spin, interacting with an infinite number of quantum oscillators whose frequencies and coupling strengths obey specific distributions. Due to its versatility, it can catch the physics of a large range of different physical systems going from defects in solids and quantum thermodynamics \cite{Lewis1988} to physical chemistry and biological systems \cite{rudnick,volker,huelga}. It has been also used to study trapped ions \cite{porras}, quantum emitters coupled to surface plasmons \cite{dzso}, quantum heat engines \cite{prx} or superconducting qubits strongly interacting with a set of individual microwave resonators that reside in a restricted frequency range \cite{PhysRevA.97.052321}. In spite of its simplicity, an exact solution is not available and a variety of approximated approaches have been adopted to investigate the ground state physical properties of the quasi-particle formed by the two-level system surrounded by the virtual excitation cloud induced by the spin-bath coupling, i.e. the spin polaron \cite{guo,chin,bera,zhou,cirac}.

In this letter we variationally address this problem, at any temperature, by following the idea of Feynman to describe the charge polaron in the Fr\"ohlich model \cite{Feynman}. The advantages provided by this method, based on the path integrals, represents a milestone in the theory of polarons. The high accuracy of the predicted free energy, polaron mass and optical properties has been confirmed by numerically exact methods \cite{devreese,defilippisprl,andrey}. Feynman, in his original paper on the Fr\"ohlich model \cite{Feynman}, first exactly eliminates the phonon degrees of freedom, by using the path integral technique. The polaron problem turns out to be equivalent to one-particle problem, described by a parabolic band within the continuum medium approximation, where a long-range, non-local in time, interaction between the electron and itself is present. The idea of Feynman is to variationally treat this equivalent one-particle problem by introducing a trial quadratic action, again non-local in time but exactly solvable, that describes approximatively the interaction of the electron with the lattice. At the best of our knowledge, so far this idea, based on the elimination of the phonon degrees of freedom followed by the variational principle, has not been applied in different contexts, maybe due to the fact that, only in a very particular case, the trial action gives rise to an exactly solvable path integral. Here we use a variational approach {\it a la} Feynman in the spin-boson model and focus our attention on the expected quantum critical transition between localized and delocalized states that occurs by increasing the coupling strength between the two-level system and the bosonic bath. After exactly eliminating the bath degrees of freedom, in analogy with the charge-polaron problem, we introduce a simple trial retarded interaction between the spin and itself simulating the true spin-bath interaction. It stems from a not exactly solvable model Hamiltonian where the spin is coupled to a finite number of harmonic oscillators whose frequencies and coupling strengths are variationally determined. The comparison with an approximation free Monte Carlo (MC) approach shows that only a very limited number of these fictitious modes is enough to correctly describe, up to very low temperatures, any physical property at the thermodynamic equilibrium. In particular, we show that the more one increases the number of the fictitious harmonic oscillators, simulating the coupling with the bath, the more one is able to correctly describe any relevant physical property at lower and lower temperatures. For the first time a variational approach at finite temperature, in the spin-boson model, is able to correctly predict: 1) in the Ohmic regime, a Beretzinski-Thouless-Kosterlitz (BTK) QPT exhibiting the typical universal jump at the critical value; 2) in the sub-Ohmic regime ($s \leq 0.5$), mean field QPT, with logarithmic corrections for $s=0.5$.

{\it The Model.} The spin-boson Hamiltonian is written as:
\begin{equation}\label{eq:definitionH}
  H=H_{Q} + H_{B} + H_{I},
\end{equation}
where: 1) $H_{Q}=-\frac{\Delta}{2}\sigma_x$ represents the free Qubit contribution: here $\Delta$ is the tunneling matrix element; 2) $H_{B}=\sum_i \omega_i a_i^\dagger a_i$ describes the bath, modelled by means of a collection of bosonic oscillators of frequencies $\omega_i$; 3) $H_{I}= \sigma_z\sum_i\lambda_i\left(a_i^\dagger+a_i\right)$ is the spin-bath interaction: $\lambda_i$ represents the coupling strength with the ith oscillator. In Eq.(\ref{eq:definitionH}), $\sigma_x$ and $\sigma_z$ are Pauli matrices with eigenvalues $1$ and $-1$, and $a_i$ and $a_i^\dagger$ denote bosonic creation and annihilation operators. The spin-bath couplings $\lambda_i$ are determined by the spectral function $J(\omega)=\sum_i\lambda_i^2\delta(\omega-\omega_i)=\frac{\alpha}{2}\omega_c^{1-s}\omega^s\Theta(\omega_c-\omega)$, where $\omega_c$ is a cutoff frequency. Here the adimensional parameter $\alpha$ measures the strength of the coupling, while the parameter $s$ distinguishes among the three different kinds of dissipation: Ohmic ($s = 1$), sub-Ohmic ($s < 1$), and super-Ohmic case ($s > 1$). We use units such that $\hbar=k_B=1$. 

{\it The Methods.} We investigate the physical features of this Hamiltonian, at thermal equilibrium, by using two different approaches. The first of them is based on the path integral technique. Here the elimination of the bath degrees of freedom leads to an effective euclidean action \cite{bulla_1,weiss}:
\begin{equation}\label{eq:eqS}
  S=\frac{1}{2}\int_0^{\beta} d\tau \int_0^{\beta} d\tau^{\prime} \sigma_z(\tau)K(\tau-\tau^{\prime})\sigma_z(\tau^{\prime}),
\end{equation}
where the kernel is expressed in terms of the spectral density $J(\omega)$ and the bath propagator: $K(\tau)=\int_0^{\infty} d\omega J(\omega) \frac{ \cosh \left[ \omega \left( \frac{\beta}{2}-\tau \right)\right]}{\sinh \left( \frac{\beta \omega}{2} \right)}$. In particular, for $\frac{1}{\omega_c}\ll\tau\ll\beta/2$, it has the behaviour: $K(\tau)\simeq\frac{\alpha(\omega_c)^{1-s}\Gamma(1+s)}{2\tau^{1+s}}$, that for $s=1$ becomes simply $K(\tau)=\frac{\alpha}{2\tau^2}$. The functional integral has to be done, with Poissonian measure, over all the possible piecewise constant functions, i.e. the world-lines $\sigma_z(\tau)$, with values $1$ and $-1$, periodic of period $\beta=1/T$, where $T$ is the system temperature. An efficient sampling of the path integral can be performed adopting a cluster algorithm \cite{rieger,bulla_1,anote}, based on the  Swendsen \& Wang  approach \cite{swang}. The critical properties of the spin-boson model in the sub-Ohmic regime have been successfully investigated through this MC technique \cite{bulla_1}. Here we extend this kind of calculation to the Ohmic regime. We emphasize that this approach is exact from a numerical point of view and it is equivalent to the sum of all the Feynman diagrams. It will be used also as a test for the variational approach described below. 

The second method is based on the variational principle and it is strictly related to the approach introduced by Feynman within the Fr\"ohlich model \cite{Feynman}. For the sake of clarity we resume the main steps. In general, in a many-body problem, one uses the Feynman-Dyson perturbation theory \cite{fetter, Mahan} starting from the Hamiltonian without interactions, so the more the strength of the coupling increases, the more the number of Feynman diagrams to be considered makes greater. To overcome this difficulty, in the charge polaron problem, Feynman introduced, as starting point for the perturbation theory, a so smart variational action that the expansion to the first order is already enough to obtain an excellent description of the physics for arbitrary coupling strength. Here we follow this idea. We adopt the ordered operator calculus \cite{Feynman1}, i.e. the operator equivalent of the Feynman path integral \cite{dnote}. The first step is the calculation of the partition function \cite{fetter, Mahan} $Z=Tr \left[ e^{-\beta H}\right] =Z_{0} U(\beta)$, where $U(\beta)=\langle T_{\tau} e^{-\int_{0}^{\beta} d\tau^{\prime}H_{I}^{(0)}(\tau^{\prime})}\rangle_{0}$. Here $Z_{0}$ is the free partition function (related to $H_{0}=H_{Q}+H_{B}$), $T_{\tau}$ is the time ordering operator, $H_{I}^{(0)}$ represents the Hamiltonian $H_I$ in the interaction representation, and $\langle ... \rangle_{0}$ denotes the ensemble average with respect to $H_{0}$. By choosing, for the trace, the basis of $H_{0}$ (it is factorized), it is possible to exactly eliminate the bath degrees of freedom by using the Bloch-DeDominicis theorem \cite{fetter, Mahan}. The partition function becomes:
\begin{equation}\label{eq:zeta}
  Z=Z_{Q}Z_{B}\langle T_{\tau} e^{\phi}\rangle_{Q},
\end{equation}
where:
\begin{equation}\label{eq:phi}
  \phi=\frac{1}{2}\int_0^{\beta}d\tau \int_0^{\beta} d\tau^{\prime} \sigma_z^{(0)}(\tau)K(\tau-\tau^{\prime}) \sigma_z^{(0)}(\tau^{\prime}),
\end{equation}
Note that, in Eq.(\ref{eq:zeta}), the ensemble average is with respect to $H_{Q}$ and that, differently from the path-integral representation, i.e. Eq.(\ref{eq:eqS}), in Eq. (\ref{eq:phi}) spin operators and not their eigenvalues appear. So far no any approximation has been used. In order to go ahead with the calculation, one can expand the exponential and use the standard perturbation theory in the many body problem. In order to avoid to evaluate a huge number of diagrams when the coupling with the bath increases, we follow the Feynman's idea and introduce a trial Hamiltonian ($H_{tr}$), where we replace the bath, characterized by a continuum distribution of harmonic oscillators, with a discrete collection of $N$ bosonic fictitious modes:
\begin{equation}\label{eq:definitionHtrial}
  H_{tr}=H_{Q} + \sum_{i=1}^{N}\tilde{\omega}_{i} b_i^{\dagger}b_i+ \sigma_z\sum_{i=1}^N \tilde{\lambda}_i\left(b_i^\dagger+b_i\right),
\end{equation}
where the parameters $\tilde{\omega}_{i}$ and $\tilde{\lambda}_{i}$ have to be variationally determined as specified in the following. Also in this case, due to the linearity of the coupling term, we can exactly eliminate the bosonic degrees of freedom, getting:
\begin{equation}\label{eq:zetatrial}
  Z_{tr}=Z_{Q} Z_{B_{tr}} \langle T_{\tau} e^{\phi_{tr}}\rangle_{Q},
\end{equation}
where
\begin{equation}\label{eq:phitrial}
  \phi_{tr}=\frac{1}{2}\int_0^{\beta}d\tau \int_0^{\beta} d\tau^{\prime} \sigma_z^{(0)}(\tau)K_{tr}(\tau-\tau^{\prime}) \sigma_z^{(0)}(\tau^{\prime}),
\end{equation}
and $K_{tr}(\tau)$ contains the propagator of these trial modes and the coupling strengths:  $K_{tr}(\tau)=\sum_{i=1}^N \tilde{\lambda}_i^2 \frac{ \cosh \left[ \tilde{\omega}_i \left( \frac{\beta}{2}-\tau \right)\right]}{\sinh \left( \frac{\beta \tilde{\omega}_i}{2} \right)}$. Now it is straigthforward to prove that the second derivative of the function $f(x)=-T \log \langle T_{\tau}e^{\phi_{tr}+x \left(\phi-\phi_{tr} \right)}\rangle_{S}$ is negative for any value of $x$ in the range $[0,1]$ \cite{bogoliubov}. This property gives rise to the following inequality: $f(x=1)-f(x=0)\le f^{\prime}(x=0)$, i.e. an upper bound for the free energy $F=-T \log Z$: 
\begin{equation}\label{eq:freeenergy}
  F-F_B \le F_{tr}-F_{B_{tr}}-T \frac{ \langle T_{\tau} e^{\phi_{tr}} \left(\phi-\phi_{tr}  \right)\rangle_{Q}}{ \langle T_{\tau} e^{\phi_{tr}}\rangle_{Q}}.
\end{equation}
This is exactly the same inequality found by Feynman within the Fr\"ohlich model (Feynman-Jensen inequality)\cite{Feynman,Feynman2}: it is a generalization of the well known Bogoliubov inequality \cite{Feynman2}. We emphasize that, in this variational formulation, only the free energy of the model Hamiltonian and the first correction enter. The knowledge of the eigenvalues and eigenvectors of $H_{tr}$, through numerical diagonalization \cite{bnote1}, allows us to calculate the right side of Eq.(\ref{eq:freeenergy}) \cite{cnote2} and then the parameters $\tilde{\omega}_{i}$ and $\tilde{\lambda}_{i}$. Finally, by considering the partial derivative of the free energy with respect to $\Delta$, $\omega_i$, and $\lambda_i$, and replacing, at the end of the calculation, $\phi$ with $\phi_{tr}$, it is possible to obtain the average values of the three terms of the Hamiltonian, i.e. $\langle H_Q \rangle$, $\langle H_B \rangle$, and $\langle H_I \rangle$.

\begin{figure}[thb]
  \includegraphics[width=1.01\columnwidth]{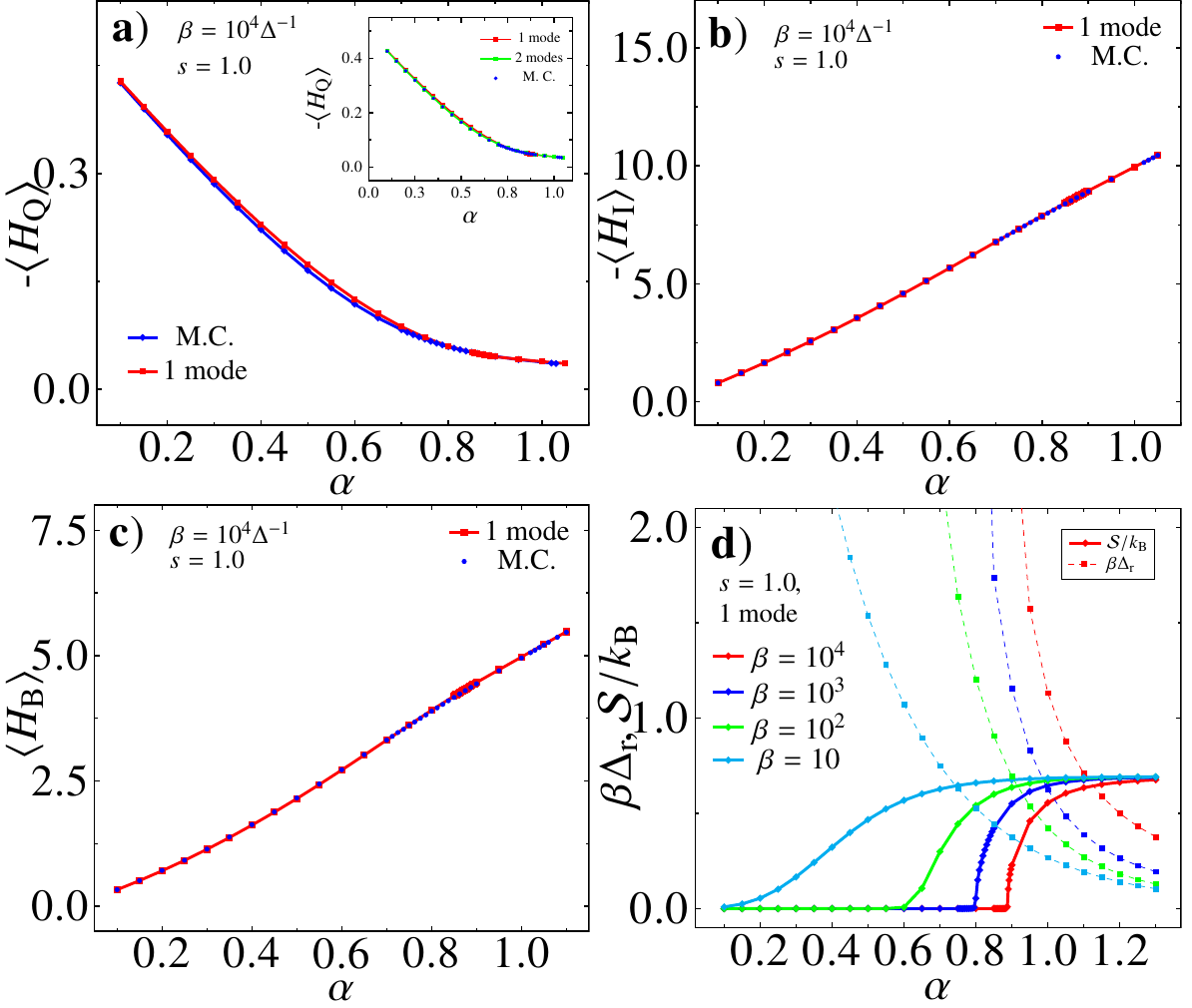}
  \caption{\label{fig:1} (color online)
    (a),(b) and (c): -$\langle H_Q\rangle$, -$\langle H_I \rangle$, and $\langle H_B\rangle$ (measured with respect to the value in the absence of interaction), in units of $\Delta$, vs $\alpha$ for $\omega_c=10 \Delta$: comparison between MC method and variational approach with one fictitious mode (in the inset $-\langle H_Q\rangle$ for $N=2$); (d) $\Delta_r/T$ (dashed line) and the entropy (solid line) vs $\alpha$. 
  }
\end{figure}

\begin{figure}[thb]
  \includegraphics[width=1.01\columnwidth]{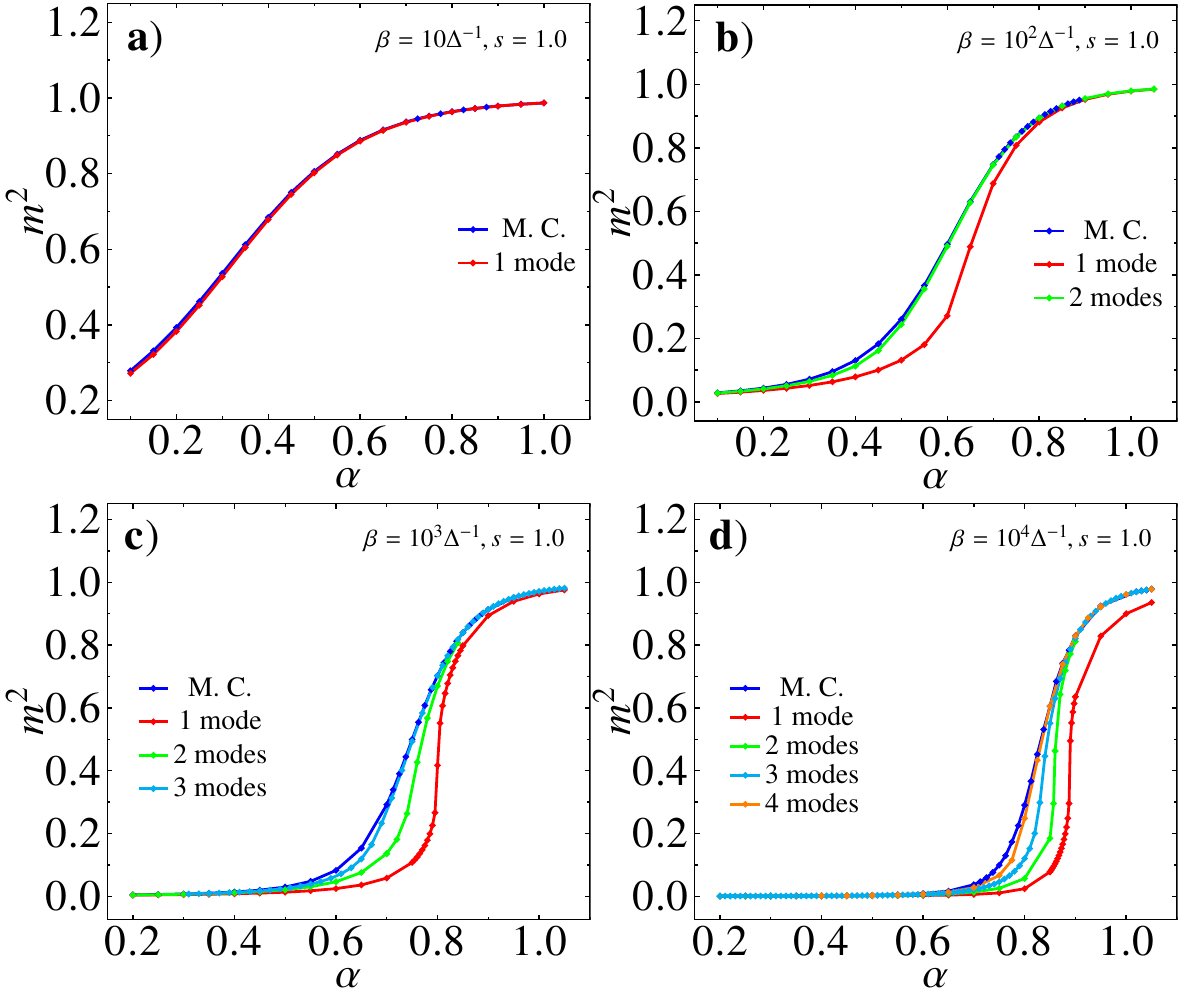}
  \caption{\label{fig:2} (color online)
    $m^2$ vs $\alpha$ for $4$ temperatures: comparison between MC method and variational approach with different fictitious modes for $\omega_c=10 \Delta$.
  }
\end{figure}

{\it The results.} We first focus our attention on the most interesting case from the physical point of view, i.e. Ohmic regime $s=1$. In Fig.~\ref{fig:1}a, Fig.~\ref{fig:1}b and Fig.~\ref{fig:1}c, we plot $-\langle H_Q\rangle$, $-\langle H_I \rangle$, and $\langle H_B\rangle$ as function of $\alpha$, at $T=10^{-4} \Delta$, by using only one fictitious mode, (N=1 in Eq.(\ref{eq:definitionHtrial})). As expected, by increasing the coupling spin-bath strength $\alpha$, $-\langle H_Q \rangle$ reduces, indicating a progressive decrease of the tunnelling between the two spin levels, whereas $\langle H_B \rangle$ and the absolute value of $\langle H_I \rangle$ increase. It is also clear that the Feynman picture provides very good estimates of the average values at thermal equilibrium independently on the value of the coupling $\alpha$. We emphasize that, for a fixed $N$, the agreement between the two approaches improves more and more by increasing the temperature. By fixing $T$, the very small differences between the two approaches practically vanish by increasing the number of fictitious modes (see inset of Fig.~\ref{fig:1}a). In Fig.~\ref{fig:1}d we plot the gap between the first two eigenvalues of the model Hamiltonian, $\Delta_r$, scaled with $T$, and the thermodynamic entropy as function of $\alpha$, for different temperatures and $N=1$. The plot points out that, at any temperature, there is a value of $\alpha$ where $\Delta_r$ becomes smaller than $T$ and, at the same time, the entropy increases from zero to $\log 2$. This behaviour suggest the presence of an incipient QPT. Actually, the existence of a QPT in the spin-boson model has been largely debated in literature \cite{emery, Anderson, leggett}. Usually the critical properties are inferred from the mapping between the spin-boson and Kondo model that is based on the well-known equivalences between Fermi and Bose operators in one dimension that, in turn, are strictly valid only in the limit of $\omega_c\rightarrow\infty$ \cite{Guinea,leggett}. It is also worth noting that the renormalization group equations of the two models are identical \cite{Chakravarty, Moore}. On the basis of these reasonings, it is generally believed that, at $s=1$, the QPT belongs to the class of BKT transitions with a critical value of $\alpha$, $\alpha_c$, around one. In order to clarify the existence of a QPT and characterize it, we study the temperature dependence of the squared magnetization $m^2=\frac{1}{\beta}\int_{0}^{\beta} d\tau \langle \sigma_z(\tau)\sigma_z(0)\rangle$. In Fig.~\ref{fig:2} we plot $m^2$ as function of $\alpha$ for different temperatures. The behaviour of $m^2$, from about $0$ to about $1$ by increasing the coupling strength, in a steeper and steeper way by lowering $T$, signals again an incipient QPT. Indeed, in the BKT transition, the quantity $m^2$ should exhibit a discontinuity, just at $\alpha_c$, for $T=0$ \cite{Kosterlitz_1973,kosterlitz1}. The plots point out that the variational approach provides an excellent agreement with the numericallly exact data of the MC method, although one has to introduce more and more modes by decreasing the temperature.

\begin{figure}[thb]
  \includegraphics[width=1.01\columnwidth]{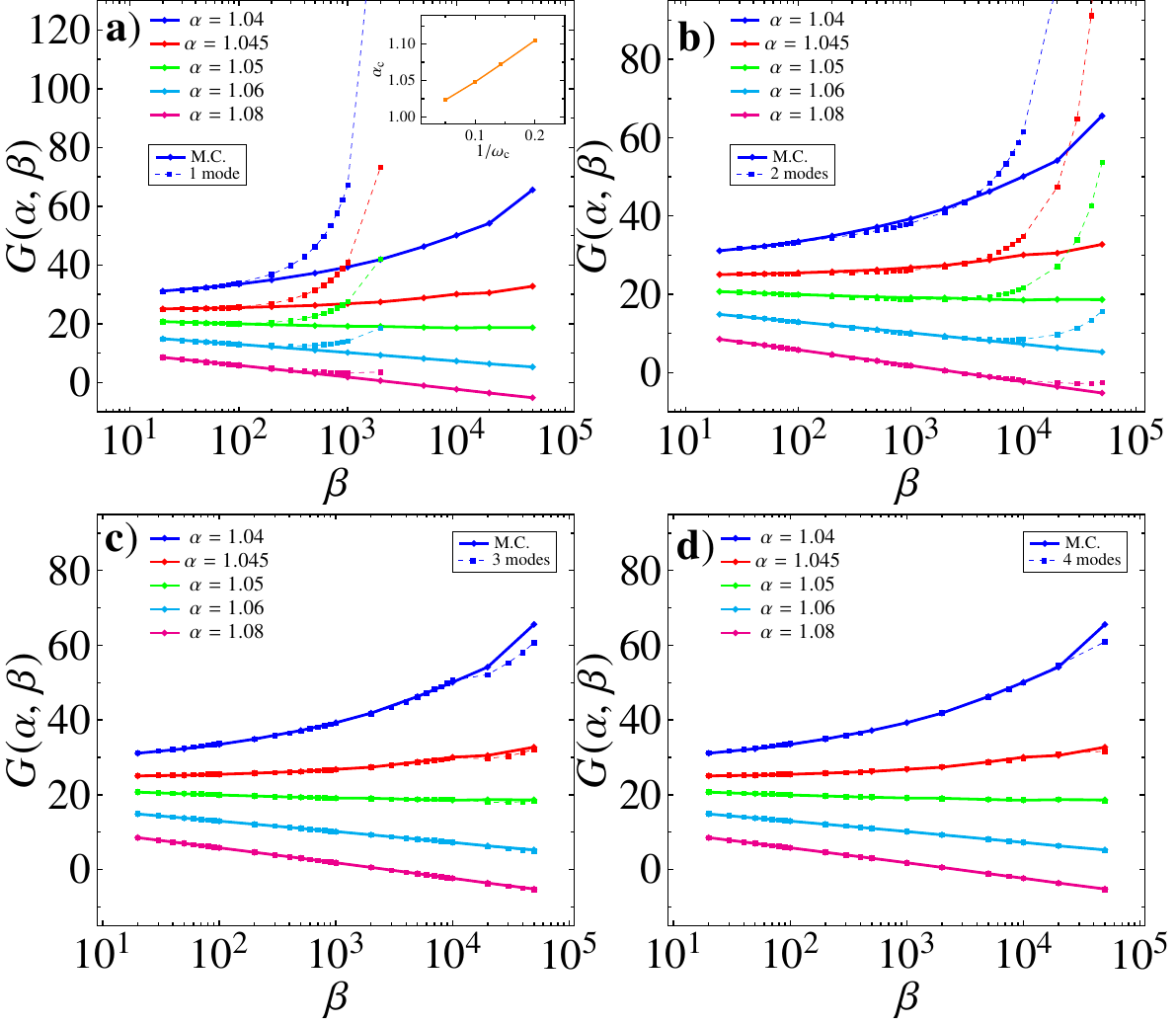}
  \caption{\label{fig:3} (color online)
    The function $G(\alpha,\beta)$ vs $\beta$ for $5$ values of $\alpha$ around $\alpha_c$: comparison between MC method and variational approach with different fictitious modes for $\omega_c=10 \Delta$. In the inset $\alpha_c$ vs $1/\omega_c$.
  }
\end{figure}

In order to get a precise estimation of $\alpha_c$, we adapt the approach suggested by Minnhagen et al. in the framework of the X-Y model, where the logarithmic behavior of the chirality as a function of the system size at the critical point is exploited \cite{minnhagen_1,minnhagen_2}. In the present context, the roles of the chirality and the lattice size are played by squared magnetization and inverse temperature $\beta$, respectively. Defining the scaled order parameter $\Psi(\alpha,\beta)=\alpha m^2$, the BKT theory predicts:
\begin{equation}
  \frac{\Psi(\alpha_c,\beta)}{\Psi_c}=1+\frac{1}{2(\ln\beta-\ln\beta_0)},
  \label{psi}
\end{equation}
where $\beta_0$ is the only fitting parameter and $\Psi_c=\Psi(\alpha_c,\beta\rightarrow\infty)$ is the universal jump that is expected to be equal to one. In this scenario, the function $G(\alpha,\beta)=\frac{1}{\Psi(\alpha,\beta)-1}-2\ln\beta$ should not show any dependence on $\beta$ at $\alpha=\alpha_c$. In order to test the validity of this scenario, in Fig.~\ref{fig:3} we plot the function $G(\alpha,\beta)$ as a function of $\ln\beta$ for different values of $\alpha$ around one. The plots clearly show that there is a value of $\alpha$ such that $G$ is independent on $\beta$. It determines $\alpha_c$, that, for $\omega_c=10 \Delta$, turns out to be about $1.05$. In the inset of Fig.~\ref{fig:3}a we plot the value of $\alpha_c$ for different values of $\omega_c$. The data point out that $\alpha_c$ is always greater than one and tends to one only when $\omega_c\rightarrow\infty$. We emphasize that $\Psi_c=1$ and $\alpha_c > 1$ imply $m^2 < 1$ for $\alpha > \alpha_c$ and $T=0$. The presence of residual tunnelling, due to a not full magnetization, explains why $\langle H_Q \rangle$, in Fig.\ref{fig:1}a, assumes, for $\alpha > \alpha_c$, a finite value different from zero.
It is also worth noting that $4$ fictitious bosonic modes, variationally determined {\it a la} Feynman, provide an execellent agreement with the numerically exact MC data, even around the critical value of the spin-bath coupling, up to $T \simeq 10^{-4}-10^{-5}\Delta$. This confirms the power of this variational approach: in this regard, it is worth mentioning that, so far, there is no any other variational formulation, based on the wavefunctions and without using the time ordered operators, able to restore the above found variational results.

\begin{figure}[thb]
  \includegraphics[width=1.01\columnwidth]{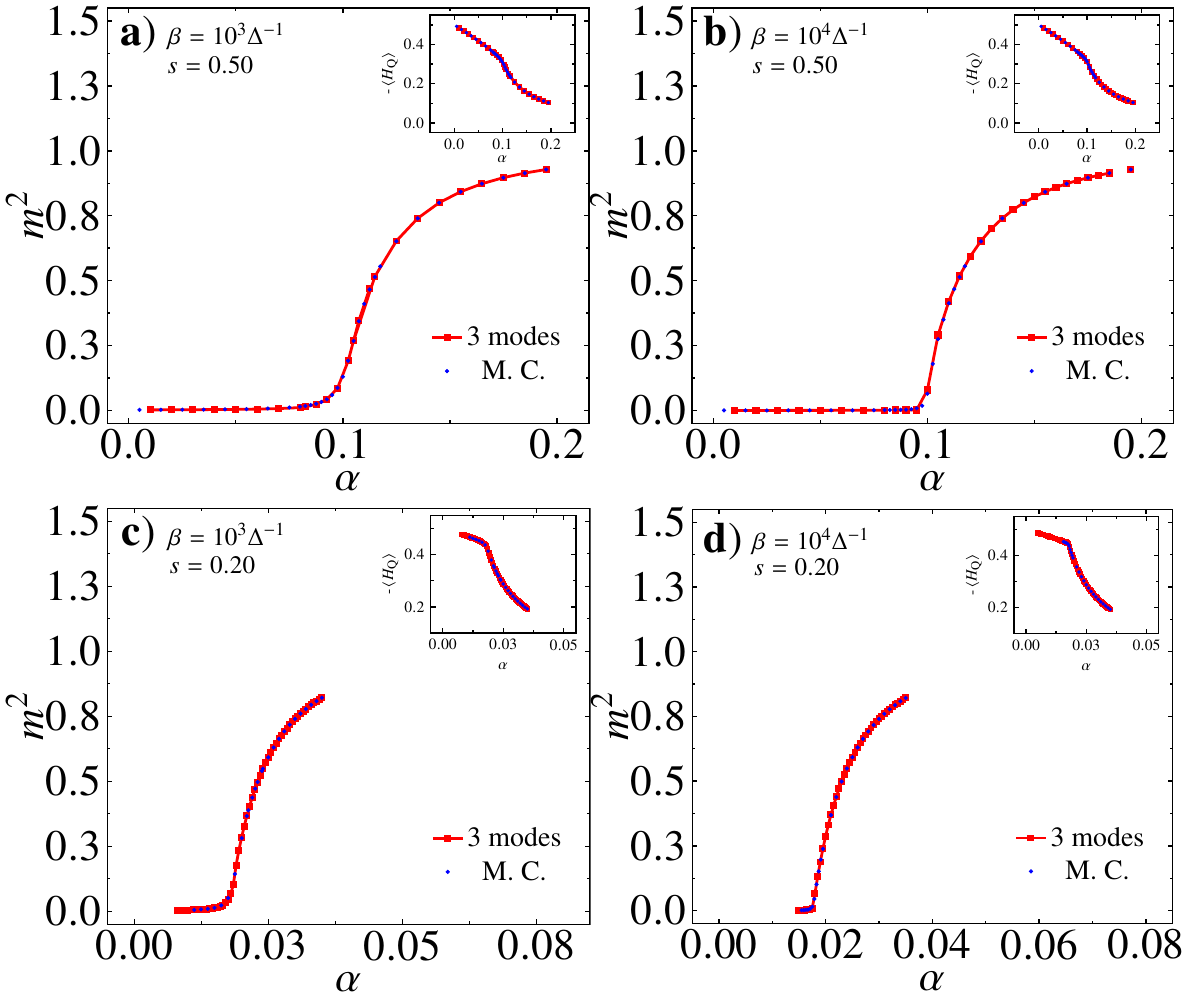}
  \caption{\label{fig:4} (color online)
    $m^2$ vs $\alpha$ for two different values of $s$: $s=0.5$ in (a) and (b), $s=0.2$ in (c) and (d). In the insets $-\langle H_Q \rangle$, in units of $\Delta$, vs $\alpha$. Comparison between MC method and variational approach {\it a la} Feynman for $\omega_c=10 \Delta$. 
  }
\end{figure}
In Fig.\ref{fig:4}, we plot $m^2$ and $-\langle H_Q \rangle$ vs $\alpha$ at two different values of $\beta$, for $s=0.5$ and $s=0.2$, i.e. sub-Ohmic regime. The variational approach {\it a la} Feynman with $3$ fictitious modes provides a remarkable agreement with the MC data, again confirming the effectiveness of this approach in any regime and for any value of the spin-bath coupling. Furthermore we emphasize that our data are in perfect agreement with the results obtained by Winter et al. \cite{bulla_1} both at $s=0.2$ and $s=0.5$. In particular, the finite $\beta$ scaling analysis for $s=0.5$ points out the presence of logarithimic corrections to the mean field theory.        

{\it Conclusion.} We investigated the spin-boson model by using two different approaches: one unbiased, exact from the numerical point of view and based on the worldline MC technique, and the other one variational and based on the method proposed by Feynamn to treat the charge polaron problem. We proved that this variational approach is extremely powerful and efficient  providing an excellent description of the physical features of the spin-polaron in any regime. In particular, we confirmed that, within the Ohmic regime, a BKT quantum phase transition sets in, proving that $\alpha_c > 1$ and $\alpha_c \rightarrow 1$ when $\omega_c \rightarrow \infty$. These results can open the way to the study of the out of equilibrium properties \cite{plenio,grifoni}, such as ultrafast processes and dynamics in the presence of strong driving fields, where standard perturbation theories fail. Work in this direction is in progress.    

\bibliography{paper}{} 

\begin{thebibliography}{46}
\bibitem [{ano()}]{anote}%
  \BibitemOpen
  \href@noop {} {}\bibinfo {note} {Starting from a world-line $\sigma_z(\tau)$,
  one introduces a number (not necessarily even) of potential flips, extracted
  from a Poissonian distribution with average $\beta\Delta/2$. Given the
  segments individuated by the real and potential flips, one connects two
  segments having the same value of the spin with the probability
  $p([u_1,u_2],[u_3,u_4])=1-\exp\left[-2\int\limits_{u_1}^{u_2}\!d\tau\int\limits_{u_3}^{u_4}\!d\tau^\prime\,
  K(|\tau-\tau^\prime|)\right]$. One then flips the connected clusters with
  probability $1/2$, and finally removes the flips that do not separate
  segments with a different value of the spin.}\BibitemShut {Stop}%
\bibitem [{dno()}]{dnote}%
  \BibitemOpen
  \href@noop {} {}\bibinfo {note} {Differently from the path integral
  technique, this procedure allows to point out the crucial role of the
  eigenstates and eigenvalues of the trial hamiltonian in determining not only
  the free energy, but any property of physical interest. The ordered operator
  calculus can be effectively used when the trial action gives rise to a not
  exact resolvable path-integral, as in the case treated in the
  text.}\BibitemShut {Stop}%
\bibitem [{bno()}]{bnote1}%
  \BibitemOpen
  \href@noop {} {}\bibinfo {note} {The exact diagonalization of the model
  Hamiltonian can be implemented by using either the natural basis or, in a
  more effective way, the coherent state basis \cite{coherent}.}\BibitemShut
  {Stop}%
\bibitem [{cno()}]{cnote2}%
  \BibitemOpen
  \href@noop {} {}\bibinfo {note} {Note that $\frac{ \langle T_{\tau}
  e^{\phi_{tr}} \left(\phi-\phi_{tr} \right)\rangle_{Q}}{ \langle T_{\tau}
  e^{\phi_{tr}}\rangle_{Q}}=\langle \phi-\phi_{tr} \rangle_{H_{tr}}$ (see Ref.
  \cite{bogoliubov})}\BibitemShut {NoStop}%
\end{thebibliography}




%
%

\end{document}